\documentclass{raa}            

\usepackage{graphicx,times}             
\usepackage{natbib}
\usepackage{amssymb,amsmath}
\bibpunct{(}{)}{;}{a}{}{,}

\usepackage[a4paper=true,dvipdfm=true,pagebackref=true]{hyperref}
\hypersetup{colorlinks = true, linkcolor = green, anchorcolor = red, citecolor = blue, filecolor = red, pagecolor = red, urlcolor = red}

\begin{document}

   \title{Evolutionary Status of T Tauri Stars}

   \volnopage{Vol.0 (20xx) No.0, 000--000}      
   \setcounter{page}{1}          

   \author{O.V. Eretnova and A.E. Dudorov}
   
   \institute{Physical Department, Chelyabinsk State University,
             Chelyabinsk 454001, Russia; {\it eretnova@csu.ru}
\vs\no
   {\small Received~~2017 December 07; accepted~~2018~~April 10}}

\abstract{ The problem of the determination of T Tauri stars masses and ages using their evolutionary status is discussed.  We test of pre-main sequence evolutionary models of \cite{D'Antona+Mazzitelli+1994}, \cite{Dotter+etal+2008}, \cite{Bressan+etal+2012} and \cite{Chen+etal+2014}, \cite{Baraffe+etal+2015} using well determined observational parameters of 12 binary T Tauri stars (TTS) and 2 binary red dwarfs.
It is shown that the masses derived using the tracks of all models are in good agreement with the masses obtained from the observations of TTS with masses $M>0.7M_{\odot}$ (mean error $\varepsilon\sim10\% $). 
Low-mass stars with $M\le0.7M_{\odot}$ have significantly greater mean error: $\varepsilon\sim50\%$ for the tracks of Bressan et al. and Chen et al. and $\varepsilon\sim30\%$ for the other tracks.
The isochrones of all tested evolutionary models diverge for the stars with masses $M\le0.7M_{\odot}$. The difference increases with the mass decrease and can reach 10\% of Kelvin-Helmholtz time for stars with mass $M=0.2M_{\odot}$. The ages of most of the considered T Tauri stars are smaller than the Kelvin-Helmholtz time. This confirms their evolutionary status of pre-main sequence stars.
}

   \authorrunning{O.V. Eretnova \& A.E. Dudorov}            
   \titlerunning{Evolutionary Status of T Tauri Stars}  

   \maketitle

\section{Introduction}         
\label{sect:intro}

We have catalogued the information about 57 double-lined spectroscopic binary T Tauri stars with known orbital elements (\citealt{Dudorov+Eretnova+2016, Dudorov+Eretnova+2017}). There are 14 systems among of them with well determined masses and radii. It is possible to determine the masses and radii of other stars using their evolutionary status, i.e. from the  positions of stars on the evolutionary tracks and isochrones on the Hertzsprung-Russell diagram. 

\cite{Dudorov+Eretnova+2016, Dudorov+Eretnova+2017} used the tracks and isochrones of \cite{D'Antona+Mazzitelli+1994} for testing the evolutionary status of particular TTS. A number of papers were published last years concerning the evolutionary models of pre-main sequence (PMS) stars (\citealt{Siess+etal+1997, Siess+etal+2000, Baraffe+etal+2002, Baraffe+etal+2015, Dotter+etal+2008, Bressan+etal+2012, Chen+etal+2014} etc.). There is a problem to choose the most suitable system of tracks and isochrones for determination of the masses and ages of T Tauri  stars.

\cite{Matchieu+etal+2006} discussed the PMS stars evolutionary models of \cite{Simon+etal+1994}, \cite{Burrows+etal+1997}, \cite{D'Antona+Mazzitelli+1994}, \cite{D'Antona+Mazzitelli+1997}, \cite{ Baraffe+etal+1998}, \cite{Palla+Stahler+1999} and \cite{Siess+etal+2000}. They used 23 PMS stars with well-defined dynamical masses from the observations for testing theoretical models. They have shown that mases of massive stars are determined by evolutionary tracks with an accuracy of 20\%. The error of mass determination using the  tracks can be 50\% and more in the case of low-mass stars. 

\cite{Stassun+etal+2014} tested thirteen PMS evolutionary models on the basis of observational data about 13 eclipsing binary stars with masses 0.04 -- 4.0$M_{\odot}$. They concluded that the error does not exceed 10\% in the case of stars with masses, $M\ge1M_{\odot}$. In the opposite case, the error reaches $\sim50 - 100\%$. These authors noted the model of \cite{Dotter+etal+2008} is more appropriate for the determination of stellar masses and ages due to smaller errors.

\cite{Lacour+etal+2016}, \cite{Baraffe+etal+2015}, \cite{Gillen+etal+2014}, \cite{Stempels+etal+2008} and others also estimated  masses and ages of individual young eclipsing binary stars by evolutionary models. All authors note the problem of the correspondence between the theory and observational data for low-mass stars. Therefore, comparison of new evolutionary models of PMS stars with the observational data remains an important problem.

In this paper we compare new and modified evolutionary tracks and the isochrones of low-mass PMS stars of \cite{Bressan+etal+2012}, \cite{Chen+etal+2014} (hereafter Padova) and \cite{Baraffe+etal+2015} (hereafter BCAH15) with tracks and isochrones of \cite{D'Antona+Mazzitelli+1994} (hereafter DM94) and \cite{Dotter+etal+2008} (hereafter referred to as Dartmouth2008). We use the observational data on 12 TTS binaries and 2 binary red dwarfs with well determined masses. T Tauri type stars with masses $0.5M_{\odot}<M<2.5 M_{\odot}$ belong to spectral classes from F to M (\citealt{Herbig+1962}). Red dwarfs are the stars with spectral class M and masses $0.15M_{\odot}<M<0.5 M_{\odot}$.

The sample of binary T Tauri stars and their main parameters are discussed in the second section. Determination of the masses of T Tauri stars by evolutionary status and their comparison with reliable observational data is presented in Section 3. In Section 4 the ages of the sampling stars are determined by isochrones. Main results are summarized in Section 5.

\section{Sample of T Tauri stars}
\label{sect:sample}

The main parameters of binary T Tauri stars and red dwarfs with well-defined absolute and relative elements are presented in Table~\ref{tab1} . Eleven systems are observed as eclipsing double line spectroscopic binary stars (EB+SB2), 3 stars are visual spectroscopic binary (VB+SB2). Nine eclipsing binary stars are the same as in work \cite{Stassun+etal+2014}. We have added 2 eclipsing spectroscopic binary stars AK Sco, BM Ori and 3 visual spectroscopic binaries to them.

 \begin{table}[!ht]
\begin{center}
\caption[]{Parameters of T Tauri Stars.}\label{tab1}
\setlength{\tabcolsep}{1pt}
\small
 \begin{tabular}{ccccccc}
  \hline\noalign{\smallskip}
Star, period & $M_{12}(M_{\odot})$& $R_{12}(R_{\odot})$&$L_{12}(L_{\odot})$& $T_{12}(K)$& r, pc& Reference\\
  \hline\noalign{\smallskip}
 1& 2& 3& 4& 5& 6& 7\\
 \hline\noalign{\smallskip} 
\multicolumn{7}{c}{EB+CD2 both components are TTS}\\
\noalign{\smallskip}
RS Cha& 1.89$\pm$0.01& 2.15$\pm$0.06& 14.8$\pm$2.8& 7640$\pm$80& 97& \cite{Alecian+etal+2005}\\
$1^d.67$& 1.87$\pm$0.01& 2.36$\pm$0.06&13.5$\pm$2.6& 7230$\pm$70& $\eta$ Cha cluster& \cite{Clausen+Nordstrom+1980}\\
\noalign{\smallskip}
ASAS  & 1.375$\pm$0.010& 1.83$\pm$0.01& 2.05$\pm$0.16& 5100$\pm$100& 280$\pm$30& \cite{Stempels+etal+2008}\\
J052821+0338.5& 1.333$\pm$0.008& 1.73$\pm$0.01& 1.38$\pm$0.11& 4700$\pm$100& Orion OB1a & \\
$3^d.8729$& & & & & & \\
\noalign{\smallskip}
AK Sco& 1.35$\pm$0.07& 1.59$\pm$0.35& 4.09$\pm$1.23& 6500$\pm$100& 145$\pm$30 & \cite{Alencar+etal+2003}\\
$13^d.609453$& 1.35$\pm$0.07& 1.59$\pm$0.35& 4.09$\pm$1.23& 6500$\pm$100& Upper Sco-Cen & \\
\noalign{\smallskip}
V1642 Ori & 1.27$\pm$0.01& 1.44$\pm$0.05& 1.37$\pm$0.25& 5200$\pm$150& 325& \cite{Covino+etal+2004}\\
(RX J0529.4+0041)& 0.93$\pm$0.01& 1.35$\pm$0.05& 0.52$\pm$0.15& 4220$\pm$150& Orion OB1a& \\
$3^d.03772$& & & & & & \\
\noalign{\smallskip}
V 1174 Ori& 1.009$\pm$0.015& 1.339$\pm$0.015& 0.64$\pm$0.07& 4470$\pm$120& 419$\pm$21& \cite{Stassun+etal+2004}\\
$2^d.634727$& 0.731$\pm$0.008& 1.065$\pm$0.011& 0.17$\pm$0.02& 3600$\pm$100& Orion NC& \\
\noalign{\smallskip}
CoRoT& 0.670$\pm$0.01& 1.30$\pm$0.04& 0.30$\pm$0.05& 3750$\pm$200& 800$\pm$100& \cite{Gillen+etal+2014}\\
223992193& 0.495$\pm$0.05& 1.11$\pm$0.04& 0.17$\pm$0.03& 3500$\pm$200& NGC 2264& \\
$3^d.8745745$& & & & & & \\
\noalign{\smallskip}
Par 1802$^*$& 0.391$\pm$0.032& 1.73$\pm$0.02& 0.49$\pm$0.04& 3675$\pm$150& 440$\pm$45& \cite{Chew+etal+2012}\\
$4^d.673903$& 0.385$\pm$0.032& 1.62$\pm$0.02& 0.305$\pm$0.025& 3365$\pm$150& Orion NC & \\
\noalign{\smallskip}
JW 380$^*$& 0.262$\pm$0.025& 1.19$\pm$0.11& 0.213$\pm$0.017& 3590$\pm$120& 470& \cite{Irvin+etal+2007}\\
$5^d.29918$& 0.151$\pm$0.013& 0.89$\pm$0.10& 0.069$\pm$0.006& 3120$\pm$110& ONC & \\
\noalign{\smallskip}
\multicolumn{7}{c}{EB+CD2 only second component are TTS}\\
\noalign{\smallskip}
BM Ori& 5.9& 940& 2.1& 22000& 420& \cite{Popper+Plavec+1976} \\
$6^d.471$& 2.18& 5.9& 31.4& 9020& Orion Trapez.& \cite{Antokhina+etal+1989}\\
\noalign{\smallskip}
TY CrA& 3.16$\pm$0.02& 1.80$\pm$0.10& 67$\pm$12& 12000$\pm$500& 129$\pm$11& \cite{Casey+Mathieu+1989}\\
$2^d.88878$& 1.64$\pm$0.01& 2.08$\pm$0.14& 2.4$\pm$0.8& 4900$\pm$400& -& \\
\noalign{\smallskip}
EK Cep& 2.02$\pm$0.02& 1.58$\pm$0.02& 14.8$\pm$1.5& 9000$\pm$200& 190& \cite{Popper+1987}\\
$4^d.4277954$& 1.12$\pm$0.01& 1.32$\pm$0.02& 1.55$\pm$0.25& 5700$\pm$200& -& \cite{Claret+2006}\\
\noalign{\smallskip}
\multicolumn{7}{c}{VB+CD2 both components are TTS}\\
\noalign{\smallskip}
V773 Tau A& 1.54$\pm$0.14& 2.22$\pm$0.20& 2.56$\pm$0.35& 4900$\pm$150& 136.2$\pm$3.7& \cite{Boden+etal+2007}\\
$51^d.1033$& 1.332$\pm$0.097& 1.74$\pm$0.19& 1.37$\pm$0.15& 4740$\pm$200& Taurus-Auriga& \\ 
\noalign{\smallskip}
V397 Aur& 1.45$\pm$0.19& 1.53& 0.755$\pm$0.09& 4345$\pm$160& 145$\pm$8& \cite{Steffen+etal+2001}\\
(NTT 045251+3016)& 0.81$\pm$0.09& 1.46& 0.306$\pm$0.05& 3550$\pm$100& Taurus-Auriga&\\
$2525^d$& & & & & & \\
\noalign{\smallskip}
HD 98800 B& 0.699$\pm$0.064& 1.09$\pm$0.14& 0.330$\pm$0.075& 4200$\pm$150& 46.7$\pm$2& \cite{Boden+etal+2005}\\
$314^d.3$& 0.582$\pm$0.051& 0.85$\pm$0.11& 0.167$\pm$0.031& 4000$\pm$150& TW Hya & \\
\noalign{\smallskip}
  \noalign{\smallskip}\hline
\end{tabular}
\end{center}
\tablecomments{0.86\textwidth}{$^*$Components of Par 1802 and JW 380 are red drafts.}
\end{table}

Table 1 contains the stars and their periods (first column), masses $M_{12}$,  radii $R_{12}$, luminosities $L_{12}$ and effective temperatures $T_{12}$ of the components (second, third, fourth and fifth columns, respectively). The upper lines in each column show the parameters of the primary components, lover lines -- parameters of the secondary components. The sixth column contains the name of parent star formation region and the distance r to it. The last column shows the references. The parameters of stars are listed with errors if they are given in the referred papers. We use the nomenclature defined in the General Catalogue of Variable Stars (\citealt{Samus+etal+2017}). If different identifier of star is used in referred papers, it is shown in parentheses. If the star is not in the General Catalogue of Variable Stars, it's identifier from the referred article is indicated. 

\section{Masses of T Tauri stars}
\label{sect:masses}

\begin{figure}[!ht]
   \centering
   \includegraphics[width=13cm, angle=0]{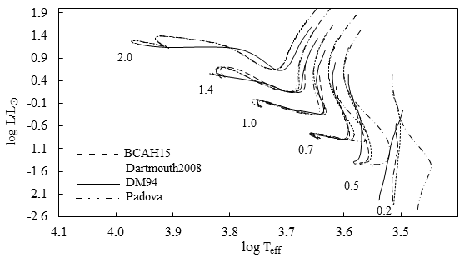}
   \caption{Evolutionary tracks for PMS stars, labeled by mass (in solar units)} 
   \label{Fig1}
   \end{figure}

Let us discuss the DM94, Dartmouth2008, Padova evolutionary models for stars with masses $0.15M_{\odot}<M<2.5 M_{\odot}$ and the BCAH15 model for stars with $0.15M_{\odot}<M<1.4 M_{\odot}$. The models are constructed for stars with chemical composition X=0.7, Y=0.28 and Z=0.02, using OPAL opacity and MLT convection theory. Figure~\ref{Fig1} and Table~\ref{tab2} show that all evolutionary tracks of stars with masses $M>0.7 M_{\odot}$ except for Padova tracks are similar a each other. The temperature steps for the Dartmouth2008 and Padova models, $\Delta{T}$=80 -- 100K, and for the BCAH15 and DM94 tracks $\Delta{T}$=150 -- 180K. It follows from Table~\ref{tab2} that the temperature difference between tracks of various models does not exceed the grid step of the tracks of given system for stars with $M>0.7 M_{\odot}$. Temperature difference for the Padova tracks is greater. In addition, the profiles of Padova tracks for small masses are very different from the other ones and have segments with a negative slope of the profile, that probably indicates the instability of  stellar models.

\begin{table}[!ht] 
\begin{center}
\caption[]{Maximum Track Offsets Relative to the Dartmouth2008 Tracks ($\Delta{T}=T-T_{Dartmouth}$)}\label{tab2}
 \begin{tabular}{ccccccccc}
 \hline\noalign{\smallskip}
& \multicolumn{8}{c} {Track masses}\\
\cline{2-9}
\noalign{\smallskip}
Model&1$M_{\odot}$& 0.9$M_{\odot}$& 0.8$M_{\odot}$& 0.7$M_{\odot}$& 0.6$M_{\odot}$& 0.5$M_{\odot}$& 0.4$M_{\odot}$& 0.2$M_{\odot}$\\
  \hline\noalign{\smallskip}
BCAH15, $\Delta{T}$(K) & +40 & +70& +60& +55& +40& +50& +45& +10\\
\noalign{\smallskip}
DM94, $\Delta{T}$(K) & +40& +70& +60& +100& +110& +150& +165& +160\\
 \noalign{\smallskip}
Padova, $\Delta{T}$(K)& +120& +210& +220& -220& -250& -250& -285& -400\\ 
  \noalign{\smallskip}\hline
\end{tabular}
\end{center}
\end{table}

\begin{figure}[!ht]
   \centering
   \includegraphics[width=13cm, angle=0]{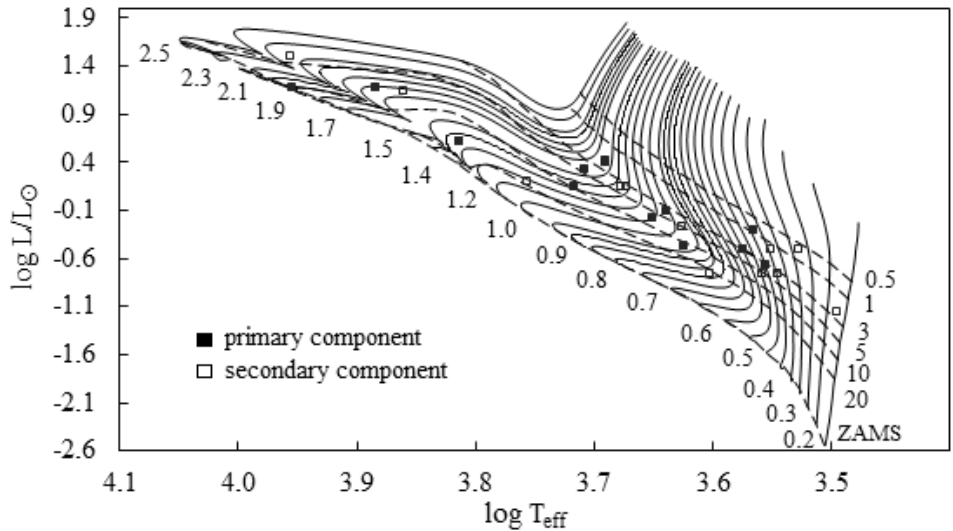}
   \caption{The Hertzsprung-Russel diagram for T Tauri stars. The Dartmouth2008 evolutionary tracks are shown by solid lines, labeled by mass (in solar units). The dashed lines are the isochrones, labeled by Myr.} 
   \label{Fig2}
   \end{figure}
   
Figure~\ref{Fig2} shows the system of the Dartmouth2008 evolutionary tracks and isochrones. The squares in the figure show the positions of binary T Tauri stars from Table~\ref{tab1} with known masses. We estimate the masses of the stars with the help of interpolation between the evolutionary tracks with errors of order of 10\%. They are presented in Table~\ref{tab3} (third, fifth, seventh and ninth columns). The top line in each column shows the mass of the primary component, and the bottom line shows the mass of the secondary component.  The fourth, sixth, eighth and tenth columns contain errors of mass estimation by tracks with respect to the observation values of the mass:
\begin{equation}
\varepsilon=\frac{M_{obs}-M_{HR}}{M_{obs}}, 
\label{eq:error}
\end{equation}
where $M_{obs}$ -- mass determined from observations, $M_{HR}$ -- mass estimated using tracks.

\begin{table}[!ht] 
\begin{center}
\caption[]{Masses of T Tauri Stars Estimated Using Tracks}\label{tab3}
\setlength{\tabcolsep}{1pt}
\small
 \begin{tabular}{cccccccccc}
 \hline\noalign{\smallskip}
Star& $M_{12obs}$& \multicolumn{2}{c} {DM94} & \multicolumn{2}{c} {Dartmouth2008}&\multicolumn{2}{c} {Padova}&\multicolumn{2}{c} {BCAH15}\\
\cline{3-10}
\noalign{\smallskip} 
& $(M_{\odot})$& $M_{12}(M_{\odot})$& $\varepsilon$& $M_{12}(M_{\odot})$& $\varepsilon$& $M_{12}(M_{\odot})$& $\varepsilon$  & $M_{12}(M_{\odot})$& $\varepsilon$\\
 \hline\noalign{\smallskip}
1& 2& 3& 4& 5& 6& 7& 8& 9& 10\\
  \hline\noalign{\smallskip}
\multicolumn{10}{c}{EB+CD2 both components are TTS}\\
\noalign{\smallskip}
RS Cha& 1.89& 2.0& 0.069& 1.79& 0.053& 1.79& 0.053& -&-\\
& 1.87& 2.0& 7.0& 1.78& 4.8& 1.77& 5.4& -&-\\
\noalign{\smallskip}
ASAS& 1.375& 1.57& -0.142& 1.58& -0.149& 1.58& -0.149& 1.50& -0.091\\
J052821+0338.5& 1.329& 1.33& 0.001& 1.31& 0.014& 1.27& 0.044& 1.38& 0.038\\
\noalign{\smallskip}
AK Sco& 1.35& 1.46& -0.082& 1.35& 0.0& 1.36& -0.007& 1.38& -0.022\\
& 1.35& 1.46& -0.082& 1.35& 0.0& 1.36& -0.007& 1.38& -0.022\\
\noalign{\smallskip}
V1642 Ori & 1.27& 1.30& -0.024& 1.31& -0.032& 1.31& -0.032& 1.29& -0.016\\
(RX J0529.4+0041)& 0.93& 0.92& 0.011& 0.95& -0.022& 0.79& 0.15& 0.92& 0.011\\
\noalign{\smallskip}
V 1174 Ori& 1.009& 1.09& -0.08& 1.10& -0.09& 1.04& -0.031& 1.10& -0.09\\
& 0.737& 0.43& 0.412& 0.50& 0.316& 0.69& 0.056& 0.44& 0.398\\
\noalign{\smallskip}
CoRoT& {\itshape0.670}& {\itshape0.53}& {\itshape0.201}& {\itshape0.550}& {\itshape0.179}& {\itshape0.70}& {\itshape-0.045}& {\itshape0.51}& {\itshape0.239}\\
223992193& {\itshape0.495}&{\itshape0.37}& {\itshape0.253}& {\itshape0.39}& {\itshape0.212}& {\itshape0.60}& {\itshape-0.212}& {\itshape0.37}& {\itshape0.253}\\
\noalign{\smallskip}
Par 1802& {\itshape0.39}1&{\itshape0.45}& {\itshape-0.151}& {\itshape0.43}& {\itshape-0.10}& {\itshape0.52}& {\itshape-0.33}& {\itshape0.42}& {\itshape-0.074}\\
& {\itshape0.385}& {\itshape0.28}& {\itshape0.273}& {\itshape0.28}& {\itshape0.273}& {\itshape0.35}& {\itshape0.091}& {\itshape0.27}& {\itshape0.299}\\
\noalign{\smallskip}
JW 380& {\itshape0.262}& {\itshape0.40}& {\itshape-0.527}& {\itshape0.44}& {\itshape-0.679}& {\itshape0.64}& {\itshape-1.443}& {\itshape0.40}& {\itshape-0.527}\\
& {\itshape0.151}& {\itshape0.13}& {\itshape-0.139}& {\itshape0.17}& {\itshape-0.126}& {\itshape0.36}& {\itshape-1.384}& {\itshape0.17}& {\itshape-0.126}\\
\noalign{\smallskip}
\multicolumn{10}{c}{EB+CD2 only second component are TTS}\\
\noalign{\smallskip}
BM Ori& 5.9& -& -& -& -& -& -& -& -\\
& 2.18& 2.4& -0.101& 2.20& -0.01& 2.18& 0.0&- &-\\
\noalign{\smallskip}
TY CrA & 3.16& -& -& -& -& -& -& -& -\\
& 1.64& 1.63& 0.006& 1.55& 0.055& 1.53& 0.067& -& -\\
\noalign{\smallskip}
EK Cep& 2.02& 2.0& 0.014& 1.91& 0.055& 1.90& 0.059& -& -\\
& 1.12& 1.18& -0.054& 1.18& -0.054& 1.18& -0.054& 1.14 & -0.018\\
\noalign{\smallskip}
\multicolumn{10}{c}{VB+CD2 both components are TTS}\\
\noalign{\smallskip}
V773 Tau A& 1.54& 1.64& -0.065& 1.55& -0.007& 1.53& 0.007& -& -\\
& 1.332& 1.35& -0.014& 1.35& -0.014& 1.30& 0.024& 1.40& -0.051\\
\noalign{\smallskip}
V397 Aur& 1.45& 1.00& 0.310& 1.01& 0.303& 0.88& 0.393& 1.02& 0.297\\
(NTT 045251+3016)& 0.81& 0.37& 0.543& 0.39& 0.519& 0.53& 0.346& 0.37& 0.543\\
\noalign{\smallskip}
HD 98800 B& {\itshape0.699}& {\itshape0.90}& {\itshape-0.288} &{\itshape0.92}& {\itshape-0.316}& {\itshape0.81}& {\itshape-0.159}& {\itshape0.86}& {\itshape-0.230}\\
& {\itshape0.582}& {\itshape0.73}& {\itshape-0.259}& {\itshape0.76}& {\itshape-0.310}& {\itshape0.73}& {\itshape-0.259}& {\itshape0.75}& {\itshape-0.293}\\
  \noalign{\smallskip}\hline
\end{tabular}
\end{center}
\end{table}

\begin{table}[!ht] 
\begin{center}
\caption[]{Mean Errors of Determination of the Masses Using Tracks and Standard Deviations}\label{tab4}
 \begin{tabular}{cccccccc}
  \hline\noalign{\smallskip}
& \multicolumn{3}{c}{$M\le0.7M_{\odot}$}& &\multicolumn{3}{c}{$M>0.7M_{\odot}$}\\
 \noalign{\smallskip}
\cline{2-4}\cline{6-8}
 \noalign{\smallskip}
Model& $\varepsilon_{m}$& $|\varepsilon|_{m}$& $\sigma$ & & $\varepsilon_{m}$& $|\varepsilon|_{m}$& $\sigma$\\
\noalign{\smallskip}
  \hline\noalign{\smallskip}
DM94 &  -0.08 & 0.261& 0.292& & 0.028& 0.115& 0.190\\
\noalign{\smallskip}
Dartmouth2008 &  -0.108& 0.274& 0.325& &  0.055& 0.096& 0.163\\
 \noalign{\smallskip}
Padova &  -0.477& 0.499& 0.593& &  0.054& 0.085& 0.131\\
\noalign{\smallskip}
BCAH15 &  -0.065& 0.262& 0.303& &  0.081& 0.133& 0.210\\
  \noalign{\smallskip}\hline
\end{tabular}
\end{center}
\end{table}

Mean values of errors of mass determination using tracks $\varepsilon_m$, mean values of absolute errors $|\varepsilon|_m$ and standard deviations $\sigma$ from mean values are given in Table~\ref{tab4}. 

Table~\ref{tab3} shows that masses $M_{HR}$ and $M_{obs}$ are in good agreement for most T Tauri stars with $M>0.7M_{\odot}$. The error, $\varepsilon\le15\%$ for all evolutionary models. The errors $\varepsilon\sim30-50\%$ only for components of NTT 045251+3016 and for secondary component of V1174 Ori. NTT 045251+3016 is visual spectroscopic binary star. The stars with $M\le0.7M_{\odot}$ have significantly large difference between $M_{HR}$ and $M_{obs}$ (they are italicized in Table~\ref{tab3}). The error $\varepsilon\le30\%$ for the T Tauri stars, while $\varepsilon$ can be $\sim100\%$ for red dwarfs. Such errors for individual stars cannot be explained by errors in the effective temperatures and luminosities (see Table~\ref{tab1}).  

The values of mean absolute errors and standard deviations are close to each other in all models for the T Tauri stars with $M>0.7M_{\odot}$ (see Table~\ref{tab4}). They are $|\varepsilon|_m\sim 10\%$ and $\sigma\sim15 - 20\%$. For stars with $M\le0.7M_{\odot}$ mean absolute errors and standard deviations are $\sim30\%$ in the DM94, BCAH15, Dartmouth2008 models and $|\varepsilon|_m\sim 50\%$, $\sigma\sim60\%$ in the Padova model. The masses found on Padova tracks were systematically larger than the masses obtained from observations for almost all stars with  $M\le0.7M_{\odot}$ (mean error $\varepsilon_m= – 47.7\%$).

\section{Ages of T Tauri Stars}
\label{sect:ages}

\begin{figure}[!ht]
   \centering
   \includegraphics[width=13cm, angle=0]{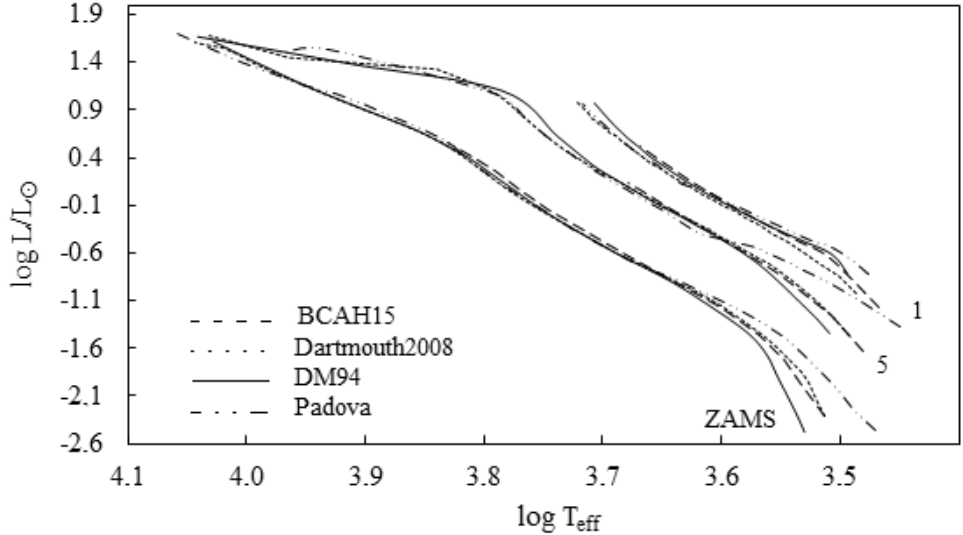}
   \caption{Isochrones for PMS stars, labeled by age (in Myr).} 
   \label{Fig3}
   \end{figure}

The position of star on the isochrones on the Hertzsprung-Russell diagram allows one to determine it's age and evolutionary status. Figure~\ref{Fig3} illustrates the isochrones of PMS stars. We estimate the ages of the stars of our sample by interpolation between the isochrones. Table~\ref{tab5} shows the ages of the stars expressed in fractions of the Kelvin-Helmholtz contraction time (\citealt{Kippenhahn+Weigert+1990}):
\begin{equation}
t_{KG}=1.6\cdot10^7\cdot\left(\frac{M}{M_{\odot}}\right)^2\cdot\left(\frac{R_{\odot}}{R}\right)\cdot\left(\frac{L_{\odot}}{L}\right),
\label{eq:time}
\end{equation}

\begin{table}
\begin{center}
\caption[]{Ages of T Tauri Stars in Fraction of the Kelvin-Helmholtz Time.}\label{tab5}
 \begin{tabular}{ccccc}
  \hline\noalign{\smallskip}
Star name& $t_{12}/t_{KG}$ & $t_{12}/t_{KG}$& $t_{12}/t_{KG}$& $t_{12}/t_{KG}$\\
& DM94& Dartmout2008& Padova& BCAH15\\
 \hline\noalign{\smallskip}
\multicolumn{5}{c}{EB+CD2 both components are TTS}\\
RS Cha& 4.44& 4.17& 3.91& -\\
& 3.95& 4.22& 3.96& -\\
\noalign{\smallskip}
ASAS& 0.62& 0.62& 0.56& 0.68\\
J052821+0338.5& 0.28& 0.29& 0.26& 0.34\\
\noalign{\smallskip}
AK Sco& 3.81& 3.81& 3.59& 3.81\\
& 3.81& 3.81& 3.59& 3.81\\
\noalign{\smallskip}
V1642 Ori & 0.76& 0.68& 0.68& 0.91\\
(RX J0529.4+0041)& 0.25& 0.26& 0.20& 0.28\\
\noalign{\smallskip}
V 1174 Ori& 0.40& 0.40& 0.32& 0.43\\
& 0.10& 0.11& 0.20& 0.11\\
\noalign{\smallskip}
CoRoT& {\itshape0.16}& {\itshape0.16}& {\itshape0.24}& {\itshape0.16}\\
223992193& {\itshape0.12}& {\itshape0.16}& {\itshape0.30}& {\itshape0.14}\\
\noalign{\smallskip}
Par 1802& {\itshape0.38}& {\itshape0.31}& {\itshape0.51}& {\itshape0.38}\\
& {\itshape0.21}& {\itshape0.16}& {\itshape0.31}& {\itshape0.18}\\
\noalign{\smallskip}
JW 380& {\itshape0.69}& {\itshape0.80}& {\itshape1.37}& {\itshape0.69}\\
& {\itshape0.40}& {\itshape0.40}& {\itshape1.45}& {\itshape0.46}\\
\noalign{\smallskip}
\multicolumn{5}{c}{EB+CD2 only second component are TTS}\\
BM Ori& -& -& -& -\\
& 12.1& 10.9& 12.1& -\\
\noalign{\smallskip}
TY CrA& 1.51& 1.51& 1.51& -\\
& 0.34& 0.29& 0.29& -\\
\noalign{\smallskip}
EK Cephei& 5.40& 3.40& 3.59& -\\
& 1.94& 1.83& 1.83& 2.14\\
\noalign{\smallskip}
\multicolumn{5}{c}{VB+CD2 both components are TTS}\\
V773 Tau A& 0.40& 0.38& 0.38& 0.45\\
& 0.38& 0.38& 0.29& 0.37\\
\noalign{\smallskip}
V397 Aur& 0.12& 0.14& 0.10& 0.14\\
(NTT 045251+3016)& 0.07& 0.06& 0.11& 0.07\\
\noalign{\smallskip}
HD 98800 B& {\itshape0.56}& {\itshape0.56}& {\itshape0.42}& {\itshape0.56}\\
& {\itshape0.53}& {\itshape0.58}& {\itshape0.53}& {\itshape0.53}\\
  \noalign{\smallskip}\hline
\end{tabular}
\end{center}
\end{table}

The upper line in Table~\ref{tab5} shows the age of primary component, the lower line is the age of secondary component. The stars with $M\le0.7M_{\odot}$ are italicized. Figure~\ref{Fig3} and Table~\ref{tab5} show that the ages of T Tauri stars in our sample, defined using the tracks from various models, slightly different from each other. The radii of stars of the same age increase with the transition from the DM94 to the Dartmouth2008 and BCAH15 models and even more when moving to the Padova model. In the region of red dwarfs, isochrones diverge. The difference increases with the mass decrease  and can reach 10\% of Kelvin-Helmholtz time for with mass $M=0.2M_{\odot}$.

The RS Cha, AK Sco stars and the secondary component of EK Cephei and BM Ori have the ratio $t/t_{KG}>1$. All these stars are at the end of the PMS evolutionary stage. Errors  of masses, luminosity, effective temperatures and radii determination could lead to an underestimation of  the Kelvin-Helmholtz time~(\ref{eq:time}) or to overestimation of the age found from the tracks.

\section{Conclusion}
\label{sect:conclusion}

We compare modified evolutionary models of Padova (\citealt{Bressan+etal+2012, Chen+etal+2014}) and BCAH15 (\citealt{Baraffe+etal+2015}) with the DM94 (\citealt{D'Antona+Mazzitelli+1994}) and Dartmouth2008 (\citealt{Dotter+etal+2008}) models using well determined observational parameters of 12 TTS binaries and 2 binary red dwarfs.  

Our study shows that the masses and ages of T Tauri stars can be determined using any of the considered evolutionary models of PMS with accuracy of about 10\%.  The temperature difference between tracks of discussed models exceeds the grid step of the tracks for stars with a mass  $M\le0.7M_{\odot}$. Temperature difference for the Padova tracks is very large and the profiles of Padova tracks very different from the other ones. The stars with $M\le0.7M_{\odot}$ have significantly greater mean values of the absolute error. It is $\varepsilon\sim30\%$ for the DM94, Dartmouth2008 and BCAH15 tracks and $\varepsilon\sim50\%$ for the Padova tracks. 

The isochrones of all tested evolutionary models diverge from the stars with a masses $M\le0.7M_{\odot}$. The ages of most of the stars in our sample are smaller than the Kelvin-Helmholtz time of stars of the corresponding mass. This confirms their evolutionary status of pre-main sequence stars.

In the future, it is necessary to further improve the theoretical models of low-mass PMS   stars, as well as to increase the number of binary stars with well defined parameters from observations. The additional theoretical work is required to improve the convection theory and to take into account the effects of magnetic field and rotation on the internal structure and evolution of low-mass stars.

\begin{acknowledgements}
We thank Sergey Khaibrakhmanov for the help with translation of the text in English.
\end{acknowledgements}

\label{lastpage}

\end{document}